\documentclass[pra,twocolumn,nofootinbib]{revtex4}
\usepackage{amssymb,amsmath}
\usepackage{epsfig}
\usepackage{graphicx}

\begin{document}

\title{Long-lived $2s$ state of muonic hydrogen: population and lifetime}
%\rtitle{Long-lived $2s$ state of muonic hydrogen: population and lifetime}
%\sodtitle{Long-lived $2s$ state of muonic hydrogen: population and lifetime}
\author{V.\,P. Popov and V.\,N. Pomerantsev}
\thanks{This work was partially supported by Russian Foundation
 for Basic Research, grant No. 06-02-17156.}
%\rauthor{V.\,P. Popov  and V.\,N. Pomerantsev}
%\sodauthor{Popov, Pomerantsev}
\affiliation{Institute of Nuclear Physics, Moscow State University, 119991 Moscow, Russia}
%\date{}{*}

\begin{abstract}{ \textit {Ab initio} study of the  density-dependent population and lifetime of
the long-lived $(\mu p)_{2s}$ and the yield of $(\mu p)_{1s}$ atoms with kinetic energy
$0.9$~keV have been performed for the first time. The direct Coulomb $2s\rightarrow 1s$
deexcitation is proved to be the dominant quenching mechanism of the $2s$ state at kinetic
energy below $2p$ threshold and explain the lifetime of the metastable $2s$ state and 
high-energy $0.9$ keV component of $(\mu p)_{1S}$ observed at low densities. The
cross sections of the elastic, Stark and Coulomb deexcitation processes have been
calculated in the close-coupling approach taking into account for the first time both the
closed channels and the threshold effects due to vacuum polarization shifts of the $ns$
states. The cross sections are used as the input data in the detailed study of the atomic
cascade kinetics. The theoretical predictions are compared with the known experimental
data at low densities. The 40\% yield of the 0.9~keV$(\mu p)_{1s}$ atoms is predicted
for liquid hydrogen density.}
\end{abstract}

\maketitle

{\bf Introduction.} Exotic hydrogen-like atoms are formed in excited states and 
further evolution of their initial  distributions in quantum  numbers and kinetic
energy is defined by the competition of the radiative and  collisional-induced
processes during  the atomic cascade.  Muonic hydrogen $(\mu^{-}p)$ is of special 
interest among the exotic atoms due to its simplest structure and possibility to
investigate a number of problems  both the exotic atom physics and bound state QED.
It is well-known,  that $2s$ state  plays a particular role in $(\mu^{-}p)$ atom due
to $(2s-2p)$  Lamb shift,  $E_L=202.0$ meV, and has no analog in hadronic
atoms  (pionic, kaonic, etc.) in which strong interaction leads to  nuclear
absorption or annihilation from this state. 

As it is expected, the measuring of the $E_\textit{L}$ in the  so-called $\mu p$ Lamb
shift experiment \cite{RP} permits  to determine the mean-square charge radius of the
proton with a relative accuracy of $10^{-3}$ and results  in a better test of
bound-state QED.  The success of this experiment depends crucially upon the 
population,$\epsilon^{\rm long}_{2s}$ (per $\mu p$ atom), and lifetime,  $\tau^{\rm
long}_{2s}$, of the $2s$ state at kinetic energies below $E_\textit{L}$.  This
fraction is called the \textit{long-lived} or \textit{metastable} fraction  of the
$2s$ state as the Stark $2s\to2p$ transitions are energetically forbidden and the
rate of two-photon transition  to the $1s$ state is negligibly small as compared with
the rate of muon decay,  $\lambda_{\mu} = 4.55 \times 10^5 s^{-1}$.  The delayed
$K_{\alpha}$ $X$-rays induced during the collisions (see \cite{X} and references
therein) can also occur but have not     been observed \cite{HA2} until now.

A high-energy component of $(\mu p)_{1s}$ with kinetic energy  $\sim 0.9$ keV was recently
discovered \cite{RPD,PPK} from the  analysis of the time-of-flight spectra (at low
gas pressures $p_{H_2}$ = 16 and 64 hPa) and attributed  to non-radiative quenching
of the long-lived $2s$ state due to the formation of the muonic molecule in a
resonant $(\mu p)_{2s}+{\rm H}_2$ collision  and subsequent Coulomb deexcitation of the
$(pp\mu)^+$ complex (see \cite{WJKF} and references therein). However, a theoretical
estimation of the quenching rate in the framework of the side path model \cite{WJKF} gives $\sim 5 \times
10^{10}$ $s^{-1}$ at liquid-hydrogen atom density (LHD=4.25 $\times 10^{22}$
atoms/cm$^{3}$) that is about an order of magnitude smaller  than the 
value $4.4^{+2.1}_{-1.8} \times 10^{11} s^{-1}$ extracted experimentally \cite{PPK}.

Contrary to this assumption, in our recent paper~\cite{diff} we suggested that
observed collisional quenching  of the metastable $2s$ state and high-energy  $(\mu
p)_{1s}$ component can be explained by the direct Coulomb deexcitation (CD) process,
\begin{equation} \label{eq. 1}
(\mu^- p)_{2s} + {\rm H} \to (\mu^- p)_{1s} + {\rm H},
\end{equation}
however, the  problem of the $(\mu p)_{2s}$ metastability  
has not been the subject of this paper.
 
In the present study we confirm our previous predictions~\cite{diff} by the
\textit{ab initio} quantum-mechanical calculations of the cross sections for the
elastic scattering, Stark transitions and CD in the framework of the improved
close-coupling approach (CCA) taking for the first time into account both the closed
channels and the threshold effects due to the vacuum polarization shifts of the $ns$
states. The calculated cross sections of the above mentioned processes have been used
in the detailed calculations of the atomic cascade kinetics. The population and
lifetime of the metastable $2s$ state, and the yield of the hot (0.9~keV) 
$(\mu p)_{1s}$ component 
at wide densities range $(10^{-9 } - 1)$ LHD are predicted. 

{\bf Improved close-coupling approach.}
 The close-coupling approach has been applied \cite{PP} by the authors to describe
the elastic scattering and Stark transitions in the exotic atom-hydrogen molecule
collisions. Recently \cite{diff,KPP,pion,PPARH}, the unified treatment of the elastic
scattering, Stark transitions and CD in the collisions of the excited exotic (muonic,
pionic and antiprotonic hydrogen) atoms with hydrogen ones  have been performed in
the framework of the CCA. Here, we briefly remind the main assumptions and outline of
the approach.

 Since $(\mu^{-} p)$ is neutral and in the low-lying states much smaller than target
molecule, the distortion of the target electron structure during collision can be
neglected and moreover the exotic atom collision  with the hydrogen molecule can be
approximately treated as the exotic atom-hydrogen atom scattering. These assumptions
are supported in particularly by a good agreement of the experimental data \cite{Bad}
and our theoretical predictions \cite{JPP} for the kinetic energy distribution of 
$\pi^{-}p$ atoms at the instant of nuclear absorption at LHD. Further, since vacuum
polarization shifts of the $ns$ states are about two order of magnitude more than
both the fine and hyperfine splittings, the muonic atom states are described by the
non-relativistic hydrogen-like wave functions with the energies of $ns$ states
shifted (due to vacuum polarization) relative to the degenerate $nl$ states ($l\neq
0$).                                                              

Thus the total wave function of the system ($\mu p + H$) is expanded in terms of the
basis states constructed as tensor products of the free muonic and hydrogen atom wave
functions and angular momentum wave function of their relative motion. In the present
study we use the basis sets in which both the open and closed channels are included.
The coupled second order differential equations are solved numerically by the
modified Numerov method. The differential and integral cross sections for individual 
$nl\to n'l'$ transitions have been calculated  for the principal quantum
number $n=2-8$ and kinetic energies that are of interest in the detailed cascade
calculations. 

According to our study, the closed channel effects are more pronounced  for the
processes in the lower states ($n=2-5$) at very low collision  energies especially
near and below $ns-np$ thresholds.  Here we produce some of our results for the
$2l\to 2l'$ (elastic and  Stark scattering) and CD $2l\to 1s$
transitions (see figs.1, 2) where the effects of the closed channels are the most
significant and extremely important for the problem of metastable $2s$ state under
consideration. All presented results below $2p$ threshold and also the rates for the
CD $2s, 2p\to 1s$ transitions above $2p$ threshold have been calculated for the first
time.

\begin{figure}[!ht]                                                            
\begin{center}   \includegraphics[width=\columnwidth,keepaspectratio]{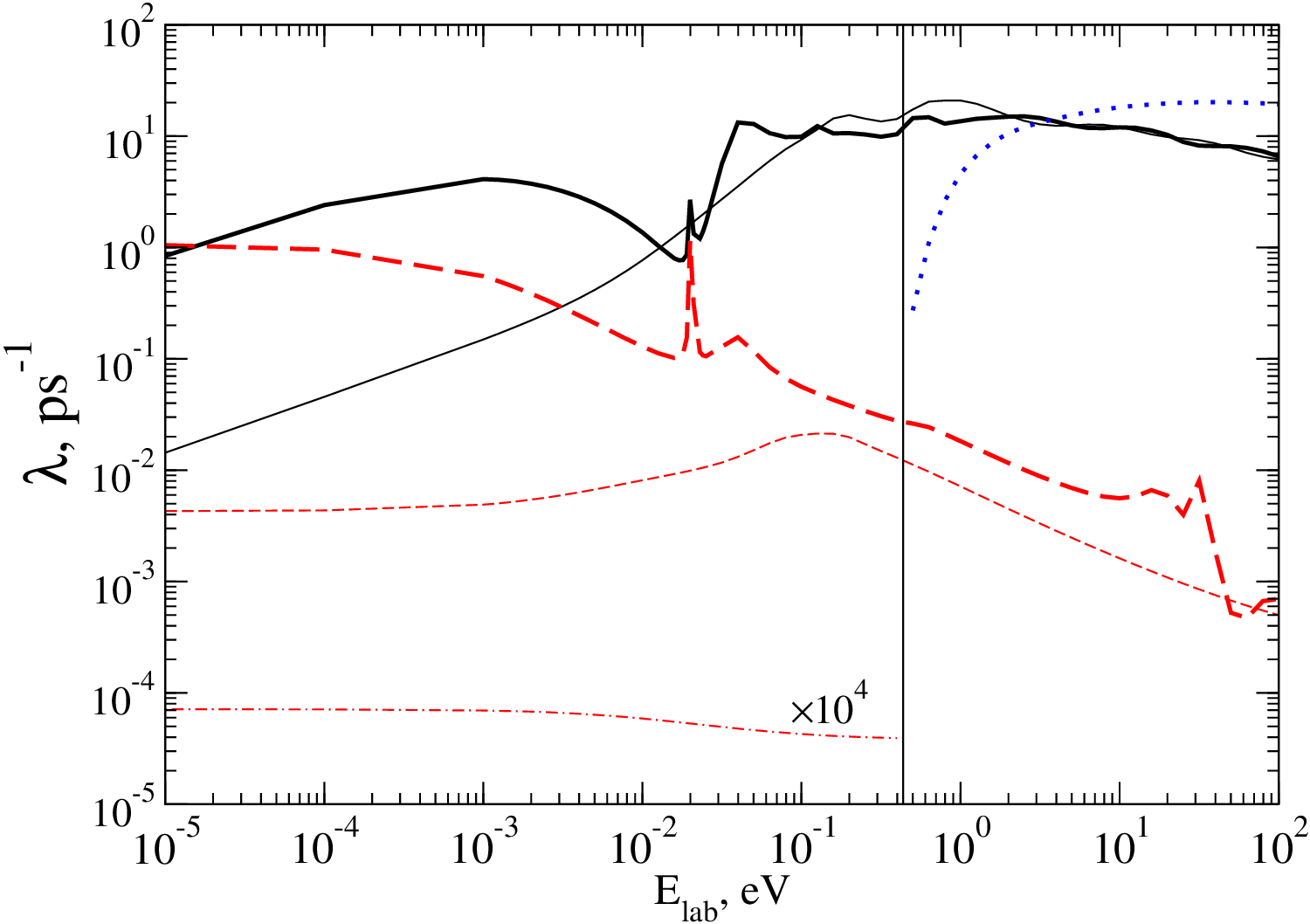}
\caption{Fig. 1. Collisional rates at LHD for the elastic $2s\to 2s$ scattering 
(solid lines), Stark $2s\to 2p$ transition (dotted line) and CD $2s\to 1s$ (dashed
lines) vs. the laboratory kinetic energy. The thin lines  show the results obtained
with the minimal basis ($n=1,2$), the thick lines -- with the extended  basis,
including all states of $(\mu^- p)$  up to $n_{\rm max}=7$. For comparison the
dash-dotted line  shows the rate of the CD $2s\to 1s$ (multiplied by $10^4$)
calculated without the closed channels. The vertical line indicates the $2s-2p$
threshold energy (0.435~eV).}  
\end{center}
\end{figure}  
\begin{figure}[!ht]                                                            
\begin{center}   \includegraphics[width=\columnwidth,keepaspectratio]{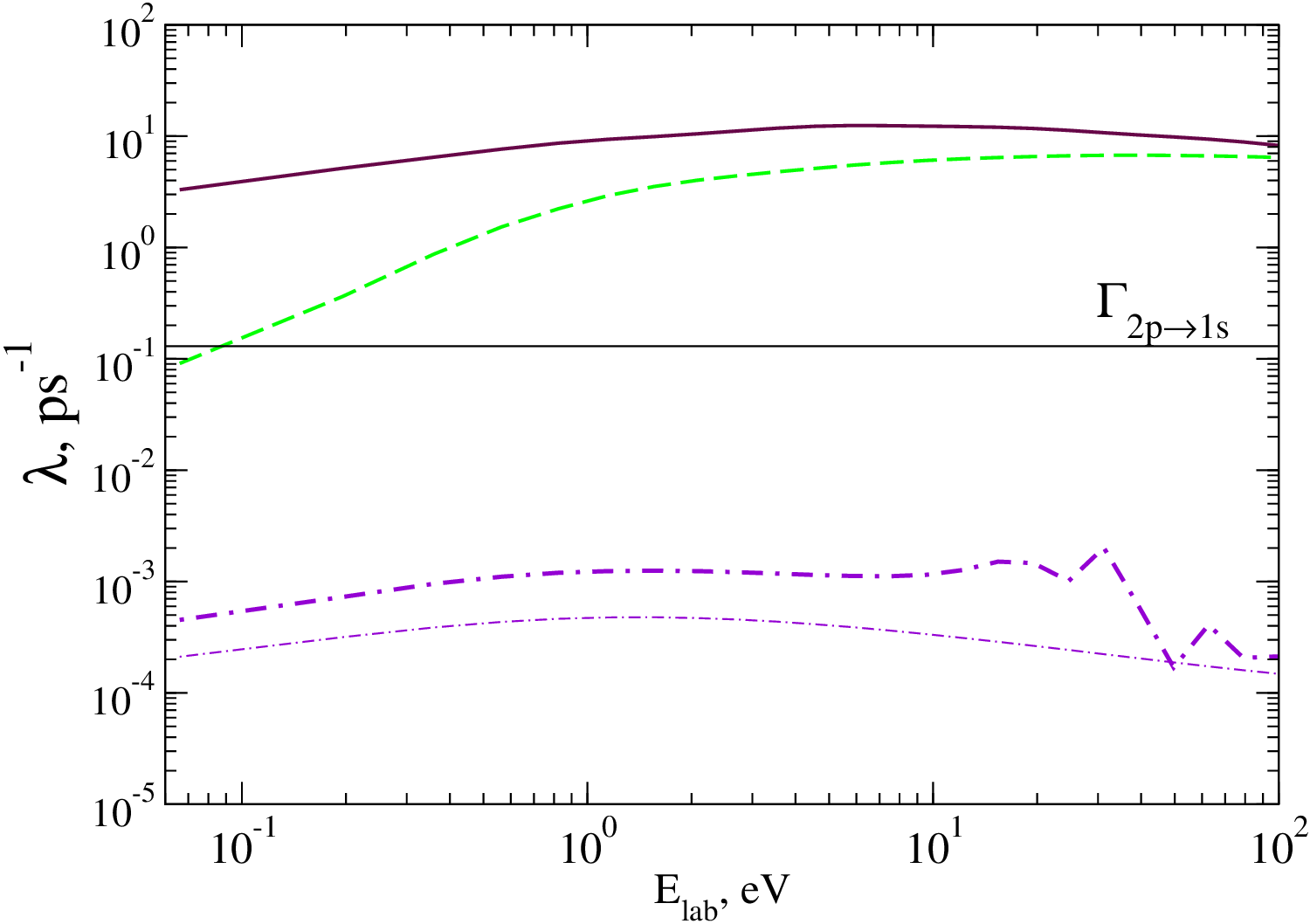} 
\caption{Fig. 2. The same as in Fig.~1 but for the elastic $2p\to2p$ (solid line), 
Stark $2p\to 2s$ ( dashed line) and CD $2p\to 1s$ (dash-dotted thin and thick lines)
processes. The energy is referred from $2p$-threshold. The radiative $2p-1s$ rate is
shown with horizontal line.}                         
\end{center} 
\end{figure}

Figures 1 and 2 show the $(\mu p)_{n=2}+H$ collisional rates, 
$\lambda_{nl\rightarrow n'l'} = LHD\sigma_{nl\rightarrow  n'l'}v,$ of the elastic and
Stark scattering and CD at LHD as a function of the laboratory kinetic energy. The
calculations have been performed for three variants of the basis set: the minimal
basis ($n=1,2$), the extended basis ($n\le n_{\rm max}=7$) and (for comparison) the
basis without closed channels at all.  At kinetic energy below  $2p$-threshold  only a few
lowest partial waves contribute to the elastic $2s\to 2s$ and CD $2s\to 1s$ cross
sections and, as it is seen from Fig. ~1, the proper description of the collisional
processes is impossible in the subspace of the open channels. The inclusion of the 
nearest closed $2p$ state into the basis leads to the tremendous increase both the
elastic $2s\to2s$ and CD $2s\to1s$ rates (about $10^3$ and $10^6$ times,
respectively).  The further extension of the basis permits to achieve the convergence
when the closed states of muonic hydrogen up to $n_{\rm max}=7$ were included. Above
$2p$ threshold  the closed channel effect is practically negligible for the elastic
and Stark scattering and is less pronounced than below $2p$ threshold for CD (see
Fig.~2). 

The results can be qualitatively explained by the different nature of the elastic/Stark
scattering and CD process. The elastic and Stark scattering is defined by the
interaction in a range about a few atomic units (a.u.) and the high partial waves
give the main contribution to the cross sections at energies above $\sim 1-2$~eV. So, the
details of the short-range interaction (at $R<0.1$~a.u.) are less important for description of these
processes excluding the scattering at very low energies. In contrast, the CD process
is accompanied by the large  energy release (tens and hundreds eV) and occurs at
substantially more small distances (about a few units  of the $\mu p$ Bohr radius,
$n^2 a_{\mu}$) so the details of the  short-range interaction and correspondingly the
closed channel effects  become more important. Besides, the number of partial waves 
involved in the CD process is always essentially smaller than in the elastic and
Stark scattering because the large centrifugal barrier for higher partial waves 
prevents the exotic atom from reaching the interaction range relevant for the CD process. 

Since the ratios between the rates of the collisional processes  are independent from
the target density, the useful observations can be derived from the results
presented in  Figs.~1,2 without the atomic cascade calculations. The CD $2p\to 1s$ is
strongly suppressed in comparison with the other cascade processes and can be
neglected in the kinetics calculations. In contrast, the rate of the CD $2s\to1s$ at
kinetic energy about a few tens meV becomes comparable with the rate of the elastic
$2s\to 2s$ scattering and therefore the CD $2s\to1s$ can occur quenching the
metastable $2s$ state. It is clear, the formation of the metastable $2s$ state can
not lead to the fast CD $2s\to1s$ transition before the muonic hydrogen is
thermalized. Finally, the comparison of the CD $2s\to1s$ rate with the one of muon
decay permits to conclude that lifetime of the metastable $2s$ state at densities
less than $\sim 10^{-6}$ LHD is mainly defined by the muon lifetime while at more
higher densities the situation is quite different and the detailed kinetics
calculations are needed. 

{\bf Atomic cascade and the initial $(n, l, E)$- distributions.}   
The cascade in the exotic atoms (see \cite{JPP, CascalII}  and references therein) is
conventionally divided into  classical (for $n \geq 9$) and quantum-mechanical  (for
$n \leq 8$) domains. In the classical domain  the results of the
classical-trajectory  Monte Carlo calculations (for details see \cite{CascalII}) of
the elastic scattering, Stark mixing and CD processes have been included with the
molecular structure of the target  taken into account.  The cross sections of the
external Auger effect have been calculated  in the semiclassical approximation
through the whole cascade.   In the quantum-mechanical domain (at $n\leq 8$)   the
differential and integral cross sections for all $nl \rightarrow n'l'$ transitions
have been calculated in the present version of the CCA. 

The initial distributions in the quantum numbers  $(n,l)$ and kinetic energy $E$ of
the exotic atom at the instant  of its formation are very important for the problem
under consideration, especially at very low target  densities relevant for the Lamb
shift experiment.  In the simplest picture of the exotic atom formation (usually used
in the cascade calculations)  the initial principal  quantum number is fixed at $n_0
\sim 14$ (for muonic hydrogen)  and the statistical $l$-distribution is assumed. 
More elaborate studies \cite{KP,KPF,JC} taking into account the molecular structure
of the target result in the initial $n$-distribution with the sharp peak at lower
values  (for muonic hydrogen at $n_0=11$) and non-statistical $l$-distribution.

     Since in the limit of the lowest densities ($\lesssim 10^{-7}$ LHD)  the atomic cascade
is mainly defined by the muon lifetime and the rates of the radiative transitions,
the information about the initial $E$- and  $(n,l)$-distributions can 
be obtained from the $K$ $X$-ray yields and the kinetic energy distribution of muonic atoms in 
$1s$ state. In the present study this knowledge was used to choose the initial distribution
parameters. We accept here the Gaussian distribution centered
at $n_0=11$ as $n$-distribution and the modified statistical $l$-distribution (at each
$n$ value): $\sim (2l+1)e^{-\alpha _l (2l+1)}$ with a one fitting parameter
 $\alpha _l$. As for the initial
kinetic energy, we use here two-exponential distribution with the parameters fitted
to reproduce the integrated kinetic energy distribution of $(\mu p)_{1s}$ extracted
\cite{RPD} from the analysis of $(\mu p)$ diffusion time at gas pressure
$p_{H_2}$=0.0625 hPa.  
   \begin{figure}[!ht] 
  \begin{center} 
  \includegraphics[width=\columnwidth,keepaspectratio]{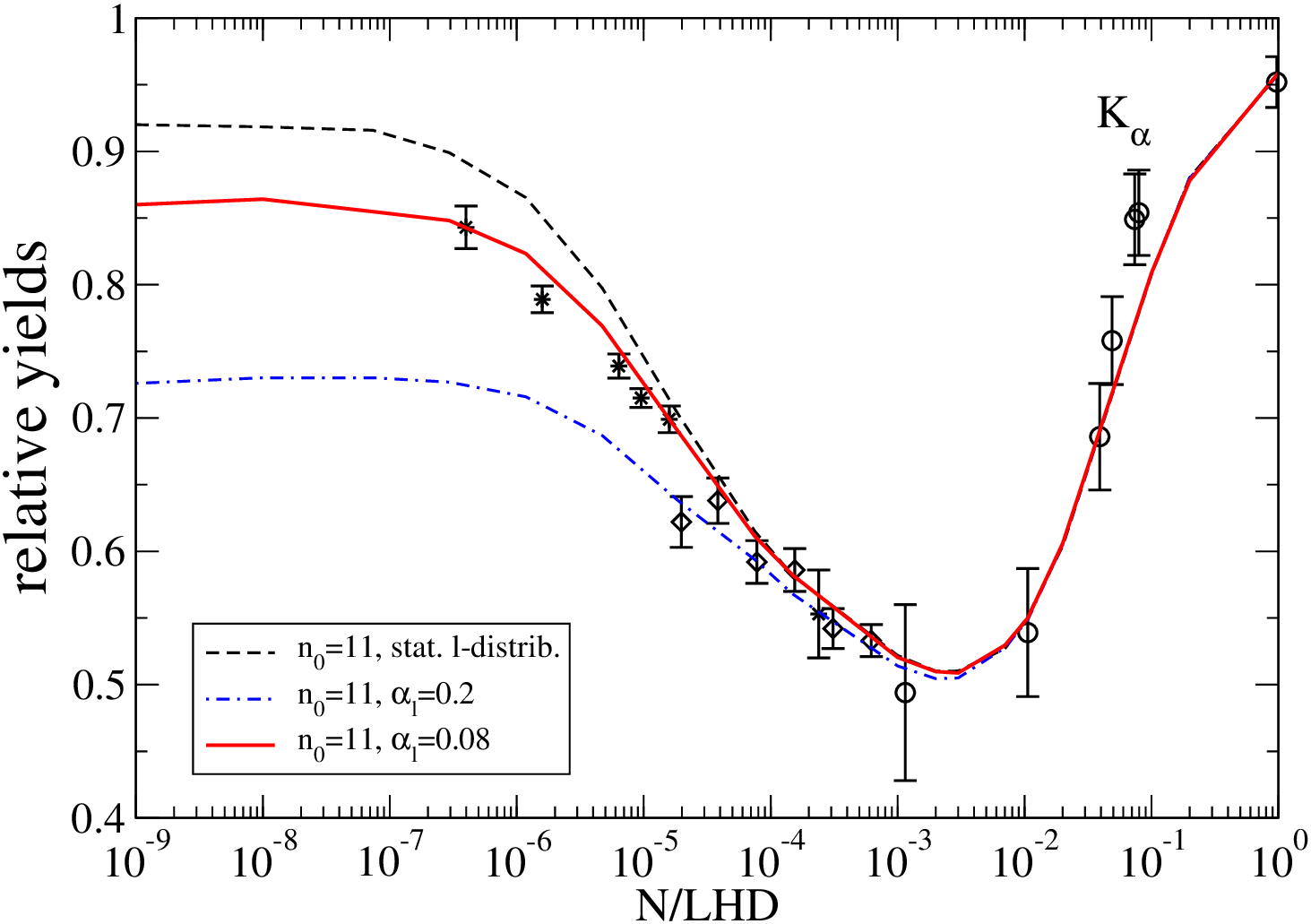}
  \caption{Fig. 3. The density dependence of the relative $K_\alpha$ 
  yield in muonic hydrogen for different variants of the initial 
  $l$-distribution: statistical (dashed line) and modified statistical 
  (solid line for $\alpha_l$=0.08 and dash-dotted line for $\alpha_l$=0.1)
  distributions. The experimental data are from \cite{HA2} (asterisks), \cite{Breg}
  (diamonds) and \cite{Laus} (open circles).}
  \end{center} 
   \end{figure}

The calculated relative $K_{\alpha}$ $X$-ray yields are shown in Fig.~3 in comparison
with the experimental data \cite{HA2, Breg, Laus}. The theoretical results illustrate
the effect of the initial $l$-distribution at fixed $n$-distribution.  It is seen
that statistical $l$-distribution results in contradiction with the experimental data
at densities $\lesssim 10^{-6}$ LHD. These differences  can not be explained by the
possible uncertainties in the rates of the collisional processes. Contrary to this,
the modified statistical distribution ($\alpha_l$=0.08) for the first time leads as a
whole to excellent agreement between theory and experimental data~\cite{HA2}. At
densities higher than $\sim 10^{-4}$ LHD the collisional transitions and Stark mixing
are more efficient and the initial $l$-distribution is practically forgotten during
the cascade. 

 {\bf Metastable $2s$ fraction: population, lifetime and yield of hot $1s$ component.} 
 The population $\epsilon ^{\rm long}_{2s}$ of the metastable $2s$ state and
the yield $\epsilon^{\rm hot}_{1s}$ of the 0.9 keV  $(\mu p)_{1s}$
atoms (due to CD $2s\rightarrow 1s$ below $2p$ threshold) calculated in the
present version of the atomic cascade are shown in Fig.~4 for two variants of the
initial $l$-distribution: statistical ($\alpha_l=0$) and modified
statistical ($\alpha_l=0.08$). 

\begin{figure}[!ht] \begin{center} 
\includegraphics[width=\columnwidth,keepaspectratio]{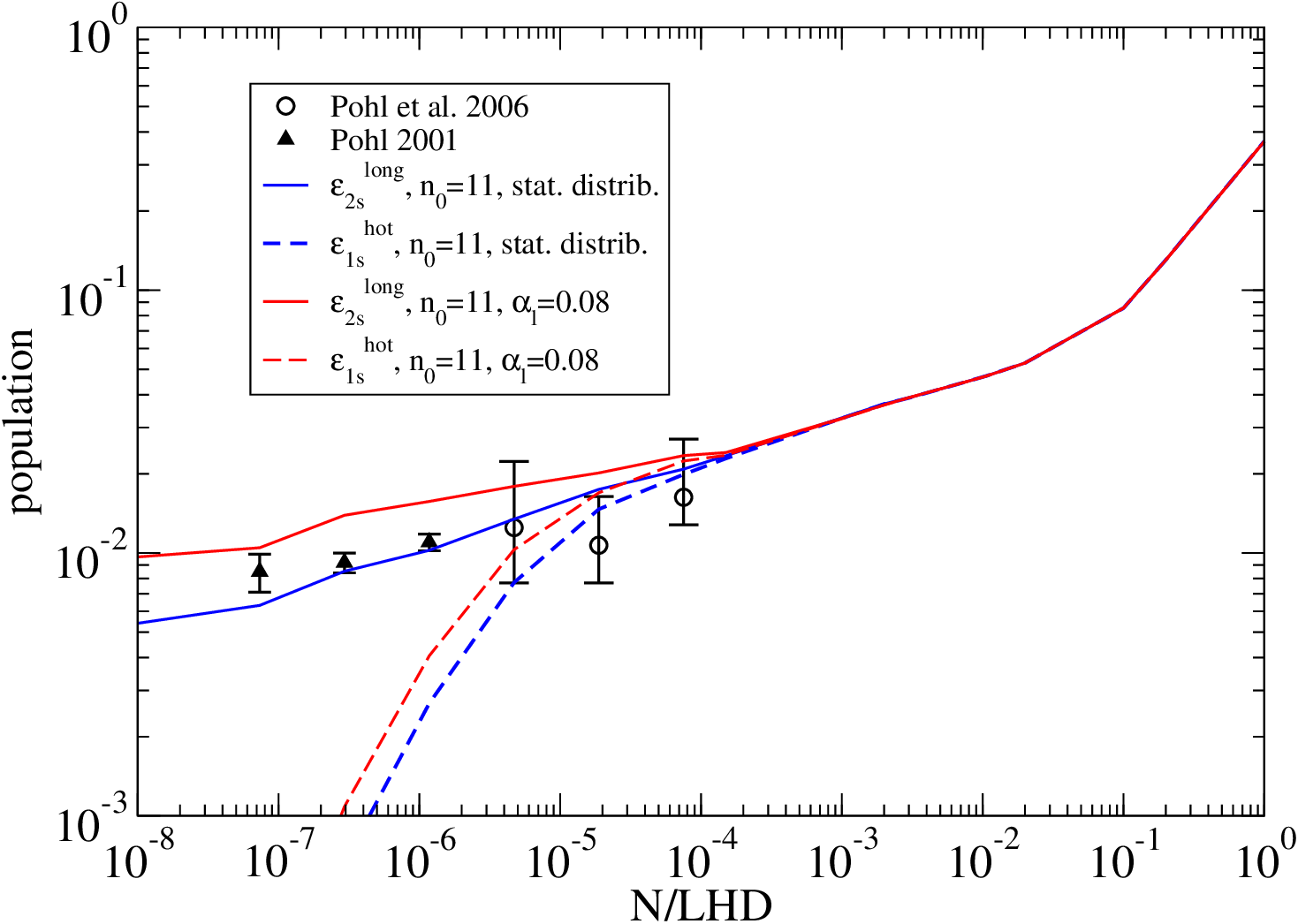}   
\caption{Fig. 4. Density dependence of the population  $\epsilon^{\rm long}_{2s}$
(solid lines) and the yield $\epsilon^{\rm hot}_{1s}$ (dashed lines) of 0.9 ~keV 
$(\mu p)_{1s}$ atoms obtained in \textit{ab initio} cascade calculations with
statistical (thin lines) and modified statistical ($\alpha_l=0.08$) (thick lines)
initial $l$-distributions. The corresponding experimental data  are from \cite{RPD}
(filled triangles) and \cite{PPK} (circles).}  
\end{center}  
\end{figure}   

The calculations with the modified $l$-distribution predict higher  populations
$\epsilon^{\rm long}_{2s}$ as compared with the ones obtained  with statistical
$l$-distribution and also with the populations derived  from the experimental data
analysis (for details see \cite{RPD,PPK}) at densities below  $10^{-4}$ LHD.
According to our study, the metastable fraction is about 1~$\%$ below $10^{-7}$ and
slowly increases in the density range $10^{-7}-10^{-2}$ LHD to $\sim 4\%$. Above 0.01
LHD the $\epsilon^{\rm long}_{2s}$ grows faster reaching $\sim 40\%$ at LHD. This
behavior and growth of the metastable fraction above 0.01 LHD is a consequence  of
the dominant role of the elastic scattering $2l\to2l$ and Stark $2p\to 2s$
transition  over the radiative $2p\to1s$ transition (see Figs.1 and 2). It is worth
noting  that density dependence of the $\epsilon^{\rm long}_{2s}$ obtained in the
present study  with the statistical $l$- distribution are similar to the one 
obtained earlier by Jensen and Markushin \cite{CascalII} only below $10^{-4}$ LHD. 
At higher densities our calculations do not confirm the sharper increase of
the population up to $65\%$ at LHD predicted in \cite{CascalII}. The theoretical
prediction for  the $\epsilon^{\rm long}_{2s}$ is in an agreement with the experimental 
result \cite{PPK} obtained in measurements of 0.9 keV $(\mu^- p)_{1s}$ at low gas
pressures $p_{H_2}$ = 16 and 64 hPa. 

The calculated for the first time yields $\epsilon^{\rm hot}_{1s}$ of the hot (0.9 keV)
fraction $(\mu p)_{1s}$ (see Fig. 4) at densities below $10^{-4}$ LHD have a
quite different density dependence as compared with the one of the $\epsilon ^{\rm
long}_{2s}$. In the density range $10^{-7}-10^{-4}$ LHD, according to our
calculations, the hot fraction increases from $0.01\%$ to $\sim 2\%$. Above $10^{-4}$
LHD the values of the $\epsilon^{\rm hot}_{1s}$ and $\epsilon ^{\rm
long}_{2s}$ is practically coincide.  Such a behavior can be explained by the growth
of the thermalized fraction of the $(\mu p)_{2s}$ atoms when the density increasing. 
At very low densities (below $10^{-7}$ LHD) the thermalized fraction of the
$(\mu p)_{2s}$ below $2p$ threshold is negligibly small and the yield of the hot
fraction is less than $0.01\%$. At higher densities it sharply increases with the
density increasing and reaches $\sim 1\%$ at $10^{-5}$ LHD.  Above $\sim10^{-4}$ LHD
the $(\mu p)_{2s}$ atoms (below $2p$ threshold) are slowed down to low energies and
then quenched in collisions by the direct $2s\to 1s$ CD and as it is demonstrated in
Fig.~4  the population $\epsilon ^{\rm long}_{2s}$ and the yield of the hot
fraction $\epsilon^{\rm hot}_{1s}$ are equal to each other. According to our study,
we predict that the yield of the 0.9 keV $(\mu p)_{1s}$ atoms and the yield of the protons 
with kinetic energy 1 keV at LHD is about $\sim 40\%$.  

\begin{figure}[!ht] 
\begin{center}
\includegraphics[width=\columnwidth,keepaspectratio]{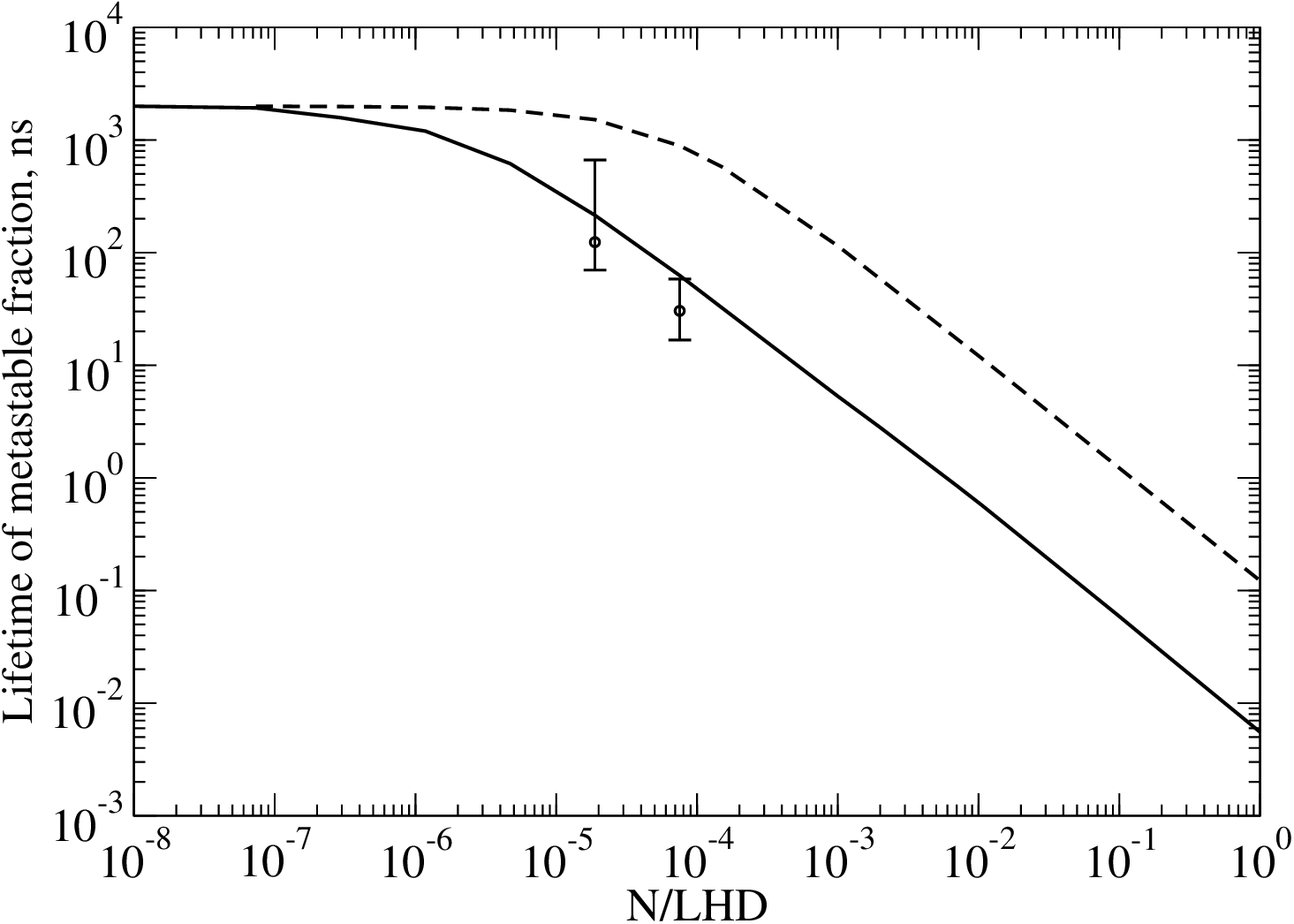}
\caption{Fig. 5. The density dependence of the lifetime, $\tau ^{\rm long}_{2s}$, 
calculated with the extended basis set with $n_{\rm max}$=7 (solid line) and with the
minimal basis set including  only closed $2p$ state (dashed line). The experimental
data are from \cite{PPK}.}
\end{center}
\end{figure}
The density dependence of the lifetime $\tau ^{\rm long}_{2s}$  calculated with the
extended basis set with  $n_{\rm max}=7$ in comparison with the result obtained with 
the basis including only the
closed $2p$ state is shown in Fig.5. The extended basis calculations are in a good 
agreement with the experimental data \cite{PPK}.
According to our study, at densities below $10^{-7}$ LHD the lifetime of the
metastable $2s$ state is mainly determined by the muon lifetime. At the density
range  $10^{-7}-10^{-5}$ LHD we observe the competition between the muon decay and
the non-radiative quenching of the metastable $2s$ state due to the direct CD 
$2s\to1s$. Above $10^{-5}-10^{-4}$ LHD the quenching of the long-lived $2s$ state is
entirely determined by the CD process.   

 {\bf Conclusion.} For the first time the cross sections for the $(\mu p)+H$
scattering process -- elastic, Stark and Coulomb deexcitation -- are calculated in
the close-coupling approach {\em including the closed channels}. These cross
sections are used as the input data for detailed study of the atomic cascade
kinetics in wide density range ($10^{-9}-1$) LHD. The results of our treatment allow
us to assert that the direct Coulomb $2s\rightarrow 1s$ deexcitation (calculated
for the first time both above and below  $2s-2p$ threshold) is the {\em dominant
quenching mechanism} of the long-lived  $2s$ fraction and explains the experimental
value of its lifetime. The calculated  population and lifetime of the metastable
$2s$-state and also predicted yield of the  hot component $(\mu p)_{1s}$ with kinetic
energy $\sim 0.9$~keV  can be very useful for the choice of the optimal conditions
for the $\mu p$ Lamb shift experiment and other experiments. In particular, it will be
very interesting to obtain experimental data about the density dependence of the
yield of the hot component $(\mu p)_{1s}$ with kinetic energy $\sim 0.9$~keV or
the yield of the protons with kinetic energy 1 keV.                                   

{\bf Acknowledgments.} We thank L. Simons for permanent interest in our studies and
fruitful discussions; T. Jensen for the fruitful cooperation in the development of the
new version of the cascade code; L. Ponomarev and the participants of the Seminar in
MUCATEX for useful discussions. This work was supported by Russian Foundation for Basic
Research (grant No. 06-02-17156).


\begin{thebibliography}{99}\itemsep -1mm
\bibitem {RP} R. Pohl, A. Antognini, F.D. Amaro et al., Can. J. Phys. {\bf 83}, 339 (2005).
\bibitem {X} L.I. Men'shikov and L.I. Ponomarev, Z. Phys. D {\bf 2}, 1 (1986).
\bibitem {HA2}H. Anderhub, H.P. von Arb, J. B\"ocklin et al., Phys. Lett. B {\bf 143}, 65 (1984). 
\bibitem {RPD} R. Pohl, Ph.D. thesis 14096, ETH Zurich, 2001.
\bibitem {PPK} R. Pohl, H. Daniel, F.J. Hartmann et al., PRL {\bf 97}, 193402 (2006).
\bibitem {WJKF} J. Wallenius, S. Jonsell, Y. Kino and P. Froelich, 
        Hyperfine Interact. {\bf 138}, 285 (2001). 
\bibitem {diff} V.N. Pomerantsev and V.P. Popov, JETP Lett. {\bf 83}, 331 (2006).
\bibitem {PP} V.P. Popov and V.N. Pomerantsev, Hyperfine Interact. {\bf 138}, 109 (2001).
\bibitem {KPP} G.Ya. Korenman, V.N. Pomerantsev, and V.P. Popov,
        JETP Lett. {\bf 81}, 543 (2005); nucl-th/0501036.
\bibitem {pion} V.N. Pomerantsev and V.P. Popov, Phys.~Rev. A {\bf 73}, 040501(R) (2006).
\bibitem {PPARH} V.P. Popov and V.N. Pomerantsev, arXiv:0712.3111.       
\bibitem {Bad} A. Badertscher, M. Daum, P.F.A. Goudsmit et al., 
        Europhys. Lett. {\bf 54}, 313 (2001).
\bibitem {JPP} T.S. Jensen, V.P. Popov and V.P. Pomerantsev, arXiv:0712.3010.
\bibitem {CascalII} T.S. Jensen and V.E. Markushin, Eur. Phys. J. D {\bf 21}, 271 (2002).
\bibitem {KP} G.Ya. Korenman and V.P. Popov, Muon Catalyzed Fusion {\bf 4}, 145 (1989). 
\bibitem {KPF} G.Ya. Korenman, V.P. Popov and G.A. Fesenko, Muon Catalyzed Fusion 
        {\bf 7}, 179 (1992).
\bibitem {JC} J.S. Cohen, Rep.Prog. Phys. {\bf 67} (2004) 1769.            
\bibitem {Breg} N. Bregant, D. Chatellard, J.P. Egger et al., Phys. Lett. A {\bf 241}, 344 (1998).
\bibitem {Laus} B. Lauss, P. Ackebauer, W.H. Breunlich et al., Phys. Rev. Lett. {\bf 80}, 3041 (1998).
\end{thebibliography}
\end{document}